\documentclass[12pt,a4]{article}
\usepackage{latexsym,epsfig,amssymb,euscript,amsmath,verbatim,cite}
\newcommand{\be}{\begin{equation}}
\newcommand{\ee}{\end{equation}}
\newcommand{\bea}{\begin{eqnarray}}
\newcommand{\eea}{\end{eqnarray}}

\numberwithin{equation}{section}
\usepackage{bm}

\begin{document}
\pagestyle{empty}
\vspace{1.8cm}

\begin{center}
{\LARGE{\bf Competition/Enhancement of Two Probe}}
{\LARGE{\bf Order Parameters in the Unbalanced}}\vspace{.2cm}
{\LARGE{\bf Holographic Superconductor}}

\vspace{1cm}

{\large{Daniele Musso\footnote{\tt dmusso@ulb.ac.be} 
\\[1cm]}}

Physique Th\'eorique et Math\'ematique\\
Universit\'e Libre de Bruxelles, C.P. 231, 1050 Bruxelles, Belgium\\

\vspace{1cm}

{\bf Abstract}
\end{center}

\noindent
We introduce and study a simple unbalanced holographic superconductor model   
with two scalar order parameters. The attention is focused on the possibility of coexisting
orderings corresponding to the concomitant condensation of two scalar operators. 
Through a probe analysis we show that an attractive or repulsive direct interaction between the
two bulk scalars leads respectively to competition and enhancement of the associated
condensates. The system at hand is a toy model for studying generic multiple ordering in a strongly 
coupled context and some comments are given about its applicability to the ferromagnetic
unconventional superconductors.



\newpage

\setcounter{page}{1} \pagestyle{plain} \renewcommand{\thefootnote}{\arabic{footnote}} \setcounter{footnote}{0}

\tableofcontents

\section{Introduction}

The $AdS$/CFT correspondence offers us the possibility
of studying quantum field theory \emph{holographically}, that is, in a dual perspective involving
an appropriate gravity theory on a spacetime with higher dimensionality.
According to the $AdS$/CFT conjecture, a classical gravitational theory on an asymptotically $AdS$ vacuum can be employed
to compute correlation functions 
of the dual quantum field theory thought of as ``living'' on the
conformal boundary of the asymptotic $AdS$ geometry \cite{Aharony:1999ti,Zaffaroni:2005ty}.
This is true in an appropriate dynamical regime
involving small curvature and weak coupling on the gravity side
and strong coupling in the dual quantum field theory.

Holography, as a framework to access the strongly coupled regime
of quantum field theory, has received vast attention since it
provides us with a new approach to study interesting toy models for 
real physical systems whose dynamics happens to be strongly coupled.
These strongly coupled models are significant both in condensed matter
and in QCD.

A particularly important field of application for holographic analysis 
is given by the physics associated with quantum phase transitions.
Within this context, models for high-$T_c$ superconductors have been widely studied holographically. 
Their non-standard properties (i.e. not describable within the BCS framework)
are thought of as being possibly intimately related to the collective and
strongly coupled dynamics of the system.
The first model for a holographic superconductor has been introduced in \cite{Gubser:2008px,Hartnoll:2008vx},
see \cite{Herzog:2009xv,Horowitz:2010gk} for a wider perspective.
The ultimate and ambitious aim of the holographic approach to unconventional superconductors would be to shed light
on the high-$T_c$ mechanism of superconduction or at least provide us with useful effective models. 
Indeed, already on a phenomenologically oriented
level, holographic studies offer the possibility of handling quantitatively some macroscopic
properties of strongly coupled superconducting systems,
such as the thermodynamic and transport properties.

As studied in \cite{Bigazzi:2011ak} it is possible
to extend the minimal holographic superconductor \cite{Hartnoll:2008vx,Hartnoll:2008kx} and introduce
an additional gauge field in the gravitational theory.
According to the holographic dictionary, such new field represents an effective way to account for a second
chemical potential in the boundary theory. Indeed, the extended holographic model
introduced in \cite{Bigazzi:2011ak} is meant to study the strongly coupled physics of unbalanced mixtures where
more than one (in that case two) chemical species are related to different chemical potentials%
\footnote{A seminal paper employing a holographic approach
to the study of unbalanced systems is \cite{Erdmenger:2011hp}.}.

Unbalanced mixtures are important both in the condensed matter context and in QCD.
In the latter the two bulk Abelian gauge fields correspond to two boundary global $U(1)$ symmetries 
which can describe the baryon number and isospin symmetries. Albeit in this paper
we are going to focus the attention (and the language)
on the condensed matter context, it should be kept in the back of one's mind that 
the holographic toy model under study allows for a wider range of applications.

According to the unbalanced superconductor interpretation of \cite{Bigazzi:2011ak},
the two chemical potentials are read as the sum and the difference
of the chemical potentials for spin-up and spin-down electrons.
In this sense, the first bulk gauge field is associated to a global ``electric'' symmetry
while the second bulk gauge field accounts for an effective ``spin'' symmetry of the boundary theory.
A non-vanishing difference among the chemical potentials for spin-up and spin-down electrons
is what makes the system \emph{unbalanced}.
As pointed out in \cite{Iqbal:2010eh}, in electronic systems at low energy
the spin-orbit coupling is suppressed leading to an effective decoupling
between spin and spacetime rotations. Therefore, even in a context where spatial rotations
happen to be broken, the low-energy $SU(2)$ spin symmetry is a valid approximate symmetry
of the IR system. The ``magnetic'' $U(1)$ in our holographic model can be thought of
as, for instance, the $\sigma_3$ component of the $SU(2)$ spin symmetry%
\footnote{In the present treatment the ferromagnetic-like order parameter
is effectively described with a scalar, namely, without loss of generality,
we understand the choice of a fixed spatial direction with
respect to which spins are projected. 
}.

Unbalanced systems are particularly interesting in view of studying 
mixed spin-electric (in one word ``spintronic'') transport properties at strong
coupling and the possible emergence of inhomogeneous FFLO phases where the superconducting
condensate acquires a non-trivial spatial modulation. 
FFLO phases at weak coupling were introduced and theoretically studied by Fulde, Ferrel,
Larkin and Ovchinnikov \cite{loff}. The first suggestion about the possibility of FFLO-like 
phases in holography has been advanced in \cite{Arean:2012ug}; 
further holographic studies have been carried out in \cite{Bigazzi:2011ak}
and separately in \cite{Alsup:2012ap,Alsup:2012kr}.

\subsection{Motivations for a second order parameter}

The purpose of the present paper is to generalize the minimal setup 
for an unbalanced holographic superconductor adding a second scalar field to the
model introduced in \cite{Bigazzi:2011ak}. 
A scalar operator which undergoes condensation describes the occurrence of
a non-trivial order parameter. For instance in the minimal holographic setup
the charged scalar breaks, upon condensing, the electric $U(1)$ symmetry and it is naturally
interpreted as a superconducting order parameter%
\footnote{This superconductivity claim can be supported with direct analysis
of the DC diverging conductivity below $T_c$, see for instance \cite{Musso:2012sn} for 
further details.}. 
In the presence of a second scalar we access the possibility of a more complicated
phase diagram where two orderings can occur. 
The dynamics and mutual relation between the two condensates is
one of the most interesting property to be studied;
in other words, one would like to investigate whether the two orderings compete or enhance one another.

The point is so far very general. Before making it more specific it is important 
to underline the versatility of a toy model possessing two order parameters. 
By simple changes of the parameters (e.g. the charges and masses of the two scalars)
one is able to describe different and interesting physical situations. 
Indeed, the toy model at hand could admit various physical interpretations and 
allows us to address the problem of ordering coexistence at strong coupling in a rather general fashion.

The focus of the present treatment is nevertheless on a specific case particularly
relevant for condensed matter applications: We take the first scalar $\psi$ to be charged
under the ``electric'' $U(1)$ and neutral to the ``magnetic'' $U(1)$ while for the second scalar $\lambda$
we consider the opposite charge assignment. Hence $\psi$ is meant to account for an electrically 
charged and magnetically neutral condensate, like for instance a condensate arising from
an s-wave Cooper-like pairing. On the other hand, $\lambda$ could represent electrically neutral magnetization%
\footnote{Notice that we have just introduced a ``magnetically charged'' order parameter
accounting for ferromagnetic-like ordering. One could also consider a neutral order parameter with
the aim of describing antiferromagnetic-like orderings, see \cite{Iqbal:2010eh,Basu:2010fa}.}.

A pivotal question in the experimental and theoretical research on 
high-$T_c$ superconductors has been the 
interplay between superconductivity and magnetic orderings
(see for instance \cite{Wakimoto1999} and references therein).
From the theoretical standpoint, standard effective techniques 
\emph{\'a la} Landau-Ginzburg have usually been employed
as, for instance, in the perturbative study of spin-density-waves in superconducting
cuprates \cite{DePrato:2006jx}%
\footnote{A study of a Landau-Ginzburg approach in the case of two order parameters is
analyzed in \cite{Ivanov:2009}. In addition, the paper contains a list of interesting physical 
applications of the general problem of coexistence of multiple orderings.}.
Complementarily, we adopt here an effective and minimal approach which nevertheless
describes an intrinsically strongly coupled medium which allows for coexistence of multiple
order parameters.

The general standard picture arising at weak coupling tends to discourage the concomitant
occurrence of superconducting and long-range ferromagnetic order \cite{Ginzburg1956}.
Moreover, in the weak coupling analysis, standard Fermi liquids present usually few instabilities and, 
more importantly here, the occurrence of an instability tends to disfavor further instability.
This is intuitively due to the fact that when the Fermi liquid undergoes an instability, a
mass gap is usually produced. Such mass gap tends to prevent, or at least discourage, further
instabilities \cite{Basu:2010fa}.

In a strongly correlated system however the picture might change radically.
In particular, in the panorama of high-$T_c$ superconductors 
a wealth of coexisting orderings such as magnetic orderings or striped phases
are possible.
The recent discovery of materials where itinerant 
ferromagnetism and superconductivity could be even cooperative (see \cite{Kakani:2011}
and references therein for details) is particularly interesting.

\section{The minimal holographic unbalanced superconductor}

We start introducing the holographic model for the minimal unbalanced superconductor in
$2+1$ dimensions in a rather succinct way; we refer to \cite{Bigazzi:2011ak} for any further detail.
The gravitational $3+1$ dimensional bulk action is:
\begin{equation}\label{laga}
 S = \frac{1}{2\kappa_4^2} \int dx^4 \sqrt{-g} \left[ R + \frac{6}{L^2} - \frac{1}{4} F_{ab}F^{ab} - \frac{1}{4} Y_{ab}Y^{ab}
 - m_\psi^2 \psi^\dagger \psi - |\partial \psi - i q_A A \psi|^2 \right] \ ,
\end{equation}
where we have an ordinary Abelian Higgs model with gauge field $A$ ($F=dA$) in the presence
of a negative cosmological constant $-6/L^2$ plus a second gauge 
field $B$ whose field strength is $Y=dB$.
We underline that the scalar $\psi$ is minimally coupled to $A$ and uncharged with respect to $B$.

We study the model considering the standard radial ansatz:
\begin{equation}
 ds^2 = -g(r) e^{-\chi(r)} dt^2 + \frac{r^2}{L^2} (dx^2 + dy^2) + \frac{dr^2}{g(r)}
\end{equation}
\begin{equation}\label{ansa}
 \psi= \psi(r)\, , \ \ \ \ \ A_a dx^a = \phi(r) dt\, , \ \ \ \ \ B_a dx^a = v(r) dt\ ,
\end{equation}
and solve the equations of motion descending from (\ref{laga}).
We skip all the analysis which has already been described in \cite{Bigazzi:2011ak} and focus on some
interesting results regarding the relation between the imbalance and the condensation of $\psi$.

We remind the reader that the boundary values for the gauge fields $A$ and $B$ are respectively
interpreted as the overall chemical potential $\mu$ and the chemical potential imbalance
$\delta\mu$ of the two fermion families (namely ``spin-up'' and ``spin-down'' electrons).
The thermodynamics of the finite temperature boundary field theory is mapped into
the thermodynamics of the dual black hole and vice versa \cite{Hartnoll:2009sz}.
To simplify the following expressions we henceforth consider the $AdS_{3+1}$ 
curvature radius to be equal to one, namely $L=1$; this can be done in full generality
and amounts to a choice of length unit.

\subsection{Phase diagram: the imbalance hinders the scalar condensation}%

\begin{figure}[t]
\centering
\includegraphics[width=70mm]{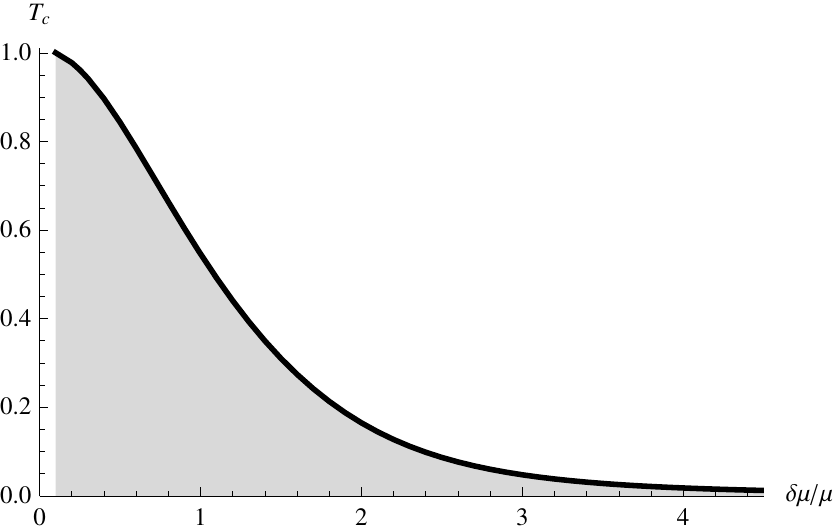} \hspace{0.5cm}
\includegraphics[width=70mm]{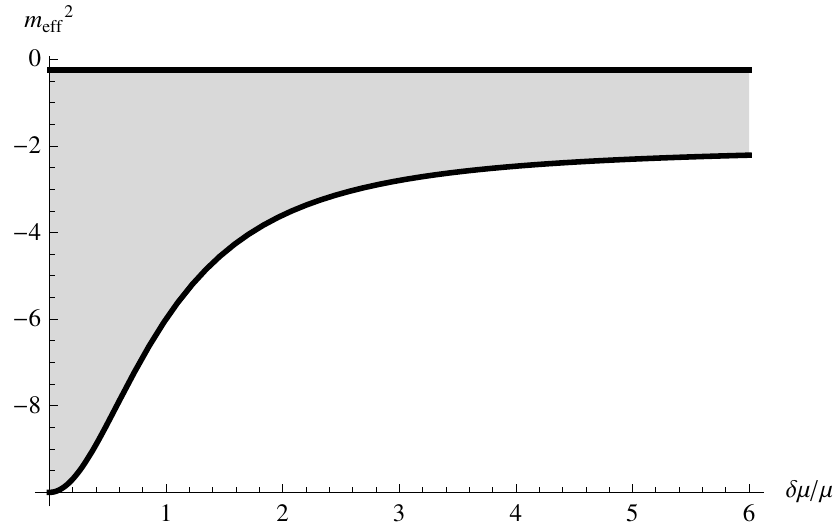}
\caption{On the left, phase diagram of the unbalanced superconductor; the shaded region corresponds 
to the superconducting condensed phase separated from the normal region by a line of second order
phase transitions. Temperature is normalized with respect to the balanced case, i.e. $T_c(\delta\mu=0)$.
On the right we have a plot showing the difference between the square of the IR effective mass and the bare mass of the bulk scalar field;
  the plot corresponds to $q=2$ and $\mu=1$.}
\label{i_vs_c}
\end{figure}

The phase diagram of the unbalanced holographic superconductor in the case of
$m_\psi^2 = -2$ is plotted in Figure \ref{i_vs_c}.
As noted in \cite{Bigazzi:2011ak} (where the diagram was obtained) it is possible to observe
that, for this specific mass value, scalar condensation appears for any value of the imbalance.
In other terms there is no Chandrasekhar-Clogston bound \cite{cc} or, equivalently, for any $\delta\mu/\mu$
there is a finite critical temperature below which the system undergoes the superconducting transition.
Note however that the bigger the imbalance, the lower the critical temperature and therefore
a larger value for $\delta\mu/\mu$ discourages condensation.
This observation is in line with the interpretation of the condensate as formed by
Cooper-like spin-singlet pairs and indeed it matches with the weak coupling expectation 
(see for instance \cite{casnar}).

The instability leading to scalar hair formation is associated to the violation of the 
Breitenlohner-Freedman bound for a scalar field 
living on the near-horizon $AdS_2$ geometry \cite{Horowitz:2009ij}.
The study of the near horizon geometry of the doubly charged Reissner-Nordstr\"{o}m
black hole solution at $T=0$ leads to the following condition for instability%
\footnote{Some details are given in Subsection 4.1 of \cite{Bigazzi:2011ak}.}
\begin{equation}\label{IR_BF}
 \tilde{m}^2_{\psi} < - \frac{3}{2}\ ,
\end{equation}
where $\tilde{m}^2$ is the effective IR mass of the scalar field given by
\begin{equation}\label{meff}
 \tilde{m}^2_{\psi} = m^2_\psi - \frac{2 q^2_A}{(1+\frac{\delta\mu^2}{\mu^2})}\ ,
\end{equation}
plotted in Figure \ref{i_vs_c}. 
Since the model we consider has $m_\psi^2=-2$, inequality (\ref{IR_BF})
is always satisfied and, as already noted, the RN doubly charged black hole is always
unstable towards hair formation at sufficiently low $T$. 
Nevertheless, let us underline that for larger values of the chemical imbalance the effective mass (\ref{meff})
is larger as well;
then, increasing $\delta\mu/\mu$ the condition for instability, though always satisfied, is met
with a smaller margin. This intuitively leads one to think that the $T=0$ doubly charged RN is ``less unstable''
when the imbalance increases and accordingly the condensation requires a lower temperature.

\section{The holographic unbalanced superconductor with two order parameters}

We introduce a second scalar field $\lambda$ in the gravitational model (\ref{laga}).
It is minimally coupled to the gauge field $B$ and neutral with respect to $A$; 
in addition, we consider a direct coupling $\gamma$ between the two scalars.
In practice we add the following terms to the action (\ref{laga}),
\begin{equation}\label{secondscalar}
 S_\lambda = \frac{1}{2\kappa_4^2} \int dx^4 \sqrt{-g} \left[ 
 - m_\lambda^2 \lambda^\dagger \lambda - |\partial \lambda - i q_B B \lambda|^2 - \gamma\ \psi^\dagger \psi\ \lambda^\dagger \lambda \right]\ .
\end{equation}
We take $m_\lambda^2=-2$ so, in the specific case at hand, $\psi$ and $\lambda$ 
have the same mass value; this value is in agreement with many instances where one studies the
$AdS_4/\text{CFT}_3$ correspondence in a UV completed framework, namely considering 
string theory consistent truncations \cite{Gauntlett:2009bh,Bobev:2011rv}.
In accordance with the ansatz (\ref{ansa}) we choose a radial
ansatz for $\lambda$ as well,
\begin{equation}\label{ansalambda}
 \lambda = \lambda(r)\ .
\end{equation}

Putting $\gamma=0$ we simply recover a ``double-copy'' of the standard superconductor 
\cite{Hartnoll:2008kx,Hartnoll:2008vx}. 
In other terms, we have two scalar/gauge vector sectors, namely $(\psi,A)$ and $(\lambda,B)$,
both minimally coupled with gravity. Let us observe that gravity couples the two sectors but,
as far as this paper is concerned, we will neglect the gravitational interaction
(i.e. we work in the probe approximation, see Section \ref{probi}) and the direct coupling $\gamma$ turns
out to be crucial%
\footnote{It would be interesting two consider instead, or in addition, also 
a direct coupling between the two gauge fields. We defer this to future work.}.

According to the ``magnetic'' interpretation of $U(1)_B$, the $\lambda$ condensate describes
a magnetic order parameter. Hence, the coexistence of the two condensates, $\psi$ and $\lambda$,
corresponds to a ferromagnetic superconducting phase%
\footnote{A treatment of ferromagnetic superconductors performed by means of Hubbard-like Hamiltonians
is described in \cite{Kakani:2011}; we refer to this paper and references therein also for a list
of interesting and recently investigated ferromagnetic superconducting systems.}. 
Without pretending to be particularly close to any specific realistic model, let us however 
observe that the ``magnetic'' degrees of freedom giving rise to a condensate resemble itinerant ferromagnetism.
Said otherwise, the degrees of freedom that we describe in terms of the ``magnetic'' condensate are
not localized or fixed to any spatial pattern or lattice; in this respect, recall that our ansatz does not present any 
specific spatial features, neither periodicity nor lattice-like structures%
\footnote{Any phenomenon of commensurate  magnetism appears then to be outside of the present treatment.
Indeed, with ``commensurate'' we mean that the spatial modulations of the order parameter correspond
to the periodicity of an underlying lattice.
However, there is an interesting interplay between possible modulated s-wave superconducting phases 
and incommensurate spin-density-wave ordering (see for instance \cite{Ptok2011}) which could be worth
studying from a holographic viewpoint.}. 
A condensate is likely to describe a plasma of degrees of freedom rather 
than an array of localized spins; this can be thought of in analogy with the fact that $\psi$ is interpreted as
a condensate of (clearly itinerant) Cooper-like pairs at strong coupling.

The total action is obtained summing (\ref{laga}) and (\ref{secondscalar}),
\begin{equation}
 S_{\text{tot}} = S + S_\lambda\ .
\end{equation}
The equations of motion descending from $S_{\text{tot}}$ are
\begin{equation}\label{phi}
 \phi'' + \left(\frac{\chi'}{2} + \frac{2}{r}\right) \phi'
 -  \frac{2 q_A^2}{g}\ \psi^2 \phi = 0
\end{equation}
\begin{equation}\label{psi}
 \psi'' + \left(\frac{g'}{g} - \frac{\chi'}{2} + \frac{2}{r}\right) \psi'
 +\frac{q_A^2 \phi^2 e^\chi}{g^2} \psi - \frac{m_\psi^2 + \gamma \lambda^2}{g} \psi = 0
\end{equation}
\begin{equation}\label{chi}
 \chi' + r \psi'^2 + r \lambda'^2 + \frac{r q_A^2 \phi^2 \psi^2 e^\chi}{g^2}
 + \frac{r q_B v^2 \lambda^2 e^\chi}{g^2} = 0
\end{equation}
\begin{equation}\label{g}
 \begin{split}
  \frac{1}{2} \psi'^2 & + \frac{1}{2} \lambda'^2 + \frac{\phi'^2 e^\chi}{4 g}
 + \frac{v'^2 e^\chi}{4 g} + \frac{g'}{g r} + \frac{1}{r^2} - \frac{3}{g}\\
&+ \frac{m_\psi^2}{2 g} \psi^2 + \frac{\gamma}{2 g} \lambda^2 \psi^2 + \frac{m_\lambda^2}{2g} \lambda^2
 + \frac{q_A^2 \psi^2 \phi^2 e^\chi}{2g^2}
 + \frac{q_B^2 \lambda^2 v^2 e^\chi}{2g^2} = 0
 \end{split}
\end{equation}
\begin{equation}\label{lambda}
 \lambda'' + \left(\frac{g'}{g} - \frac{\chi'}{2} + \frac{2}{r} \right) \lambda'
 +\frac{q_B^2 v^2 e^\chi}{g^2} \lambda - \frac{m_\lambda^2 + \gamma \psi^2}{g} \lambda = 0
\end{equation}
\begin{equation}\label{v}
 v'' + \left(\frac{\chi'}{2} + \frac{2}{r}\right)v'
 - \frac{2 q_B^2 }{g}\ \lambda^2 v = 0
\end{equation}
To solve them we consider the ansatz (\ref{ansa}) and (\ref{ansalambda}).
We study black hole solutions and the horizon is defined by $g(r_h) = 0$.
The equations of motion have the following scaling symmetry
\begin{equation}\label{scaling}
 r \rightarrow a r\ , \ \ \ \ 
 (t,x,y) \rightarrow \frac{1}{a} (t,x,y)\ , \ \ \ \ 
 g \rightarrow a^2 g \ , \ \ \ \
 \phi \rightarrow a \phi\ , \ \ \ \
 v \rightarrow a v\ .
\end{equation}
In an asymptotically $AdS_4$ background, the near boundary behavior of the gauge and scalar fields is
\begin{equation}\label{UV}
 \phi = \mu_A - \frac{\rho_A}{r} + ...  \ , \ \ 
 v = \mu_B - \frac{\rho_B}{r} + ... \ , \ \ 
 \psi = \frac{C_1}{r} + \frac{C_2}{r^2} + ... \ , \ \ 
 \lambda = \frac{D_1}{r} + \frac{D_2}{r^2} + ...
\end{equation}
To underline the generality and symmetry of the present model, we have just adopted a less specific
notation trading respectively $\mu$, $\rho$, $\delta\mu$ and $\delta\rho$ (used in the
previous Sections) for $\mu_A$, $\rho_A$, $\mu_B$ and $\rho_B$.

Note that, under the scaling (\ref{scaling}), the near-boundary quantities
behave as follows:
\begin{equation}
 \mu_{A,B} \rightarrow a \mu_{A,B}\ , \ \ \ \
 \rho_{A,B} \rightarrow a^2 \rho_{A,B}\ , \ \ \ \
 (C_1, D_1) \rightarrow a (C_1, D_1)\ , \ \ \ \
 (C_2, D_2) \rightarrow a^2 (C_1, D_1)\ .
\end{equation}
We will be concerned in studying the spontaneous condensation of the scalars and we
define the condensates as follows
\begin{equation}\label{sates}
 {\cal O}_{x} = \sqrt{2}\ C_x\ , \ \ \ \ \ \ \ 
 {\cal P}_{x} = \sqrt{2}\ D_x\ , 
\end{equation}
where $x=1,2$ and the conventional factor $\sqrt{2}$ is introduced to comply with the
existing literature; the dimension of the two kinds of condensates is $[mass]^x$.

\begin{figure}[t]
\centering
\includegraphics[width=70mm]{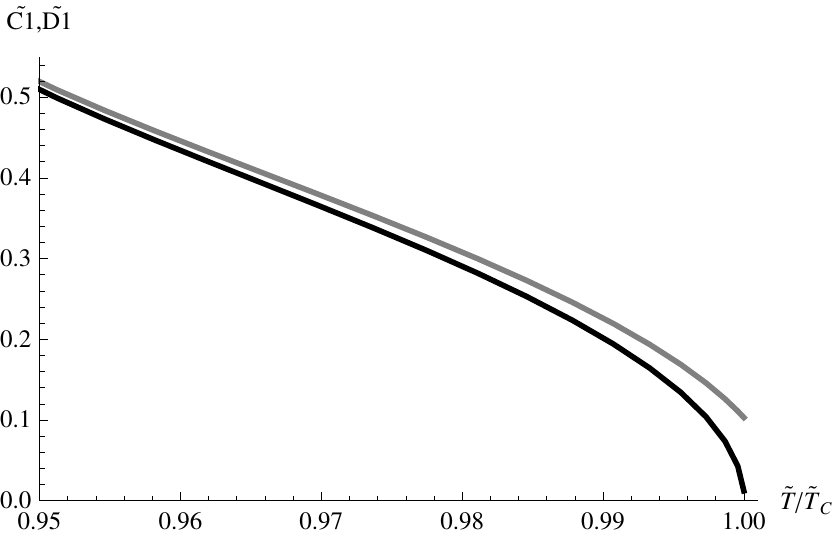} \hspace{0.5cm}
\includegraphics[width=70mm]{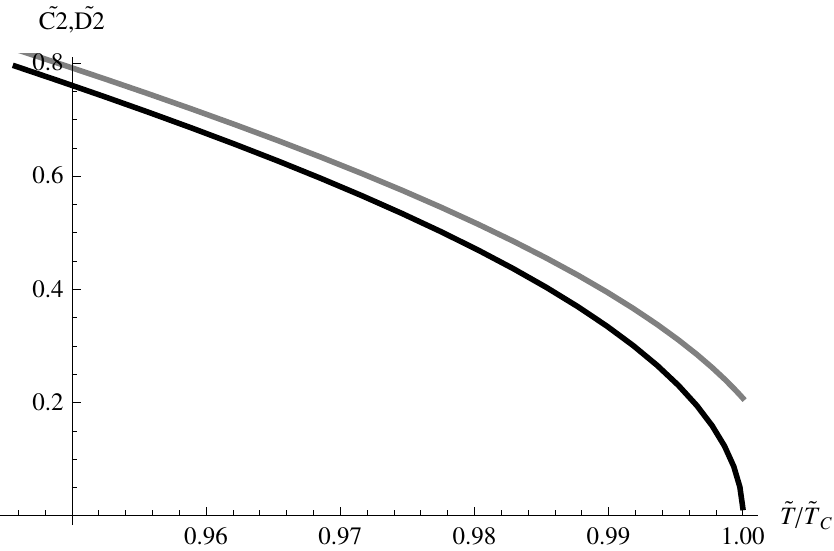}
\caption{Examples of coexisting $\psi$ (black) and $\lambda$ (gray) condensates
plotted by performing a sweep in temperature at constant charge densities. 
On the left/right plot the UV subleading/leading terms for the scalars have been respectively fixed to zero. 
The $\lambda$ (gray) plot is interrupted at a temperature value which is below its own
condensation temperature because our sweeps are technically performed ``moving''
$\psi(r_H)$, i.e. the value of $\psi$ at the horizon; indeed, we have $\psi(r_H)=0$ 
at and above the $\psi$ condensation temperature (see Section \ref{tcode}). 
The temperature is here normalized with respect to the critical 
temperature for $\psi$ condensation.}
\label{COEX}
\end{figure}

\subsection{Probe analysis}
\label{probi}

We do not have analytic solutions for the system of equations of motion, 
then we tackle the problem numerically. We work in the probe approximation that
consists in neglecting the effect of the gauge and scalar fields on the geometry.
Such approximation becomes exact in the limit of large charges $q_A,\ q_B \rightarrow \infty$
with $q_A \psi,\ q_A A,\ q_B \lambda,\ q_B B$ kept fixed and finite.
Indeed, in this limit, the terms involving the gauge and scalar fields inside
the equations of motion for the metric become negligible \cite{Hartnoll:2008kx}; 
the metric is therefore a fixed background describing an uncharged 
$AdS$-Schwarzschild black hole, namely
\begin{equation}
 g(r) = r^2 \left(1 - \frac{r_H^3}{r^3}\right) \ , \ \ \ \ \ 
 \chi(r) = 0 \ , \ \ \ \ \ 
 r_H = \frac{4}{3} \pi T \ ,
\end{equation}
where the temperature $T$ is linearly related to the black hole horizon radius $r_H$.

The equations of motion of the two gauge fields and the two scalars
are ordinary second order differential equations, namely
\begin{equation}\label{phip}
 \phi'' 
 +  \frac{2}{r}\ \phi'
 -  \frac{2 q_A^2}{g} \ \psi^2 \phi = 0\ , \ \ \ \ \ \ \
  v'' 
  + \frac{2}{r} \ v'
 - \frac{2 q_B^2 }{g}\ \lambda^2 v = 0
\end{equation}
\begin{equation}\label{psip}
 \psi'' 
 +  \left(\frac{2}{r} + \frac{g'}{g}\right) \psi'
 +\frac{q_A^2 }{g^2}\ \phi^2  \psi 
 - \frac{m_\psi^2 + \gamma \lambda^2}{g} \psi = 0
\end{equation}
\begin{equation}\label{lambdap}
 \lambda'' 
 + \left( \frac{2}{r} + \frac{g'}{g}\right) \lambda'
 +\frac{q_B^2}{g^2}\ v^2 \lambda 
 - \frac{m_\lambda^2 + \gamma \psi^2}{g} \lambda = 0
\end{equation}

Albeit the space of solutions
of the system is spanned by eight parameters, we need to impose some constraints
dictated by consistency reasons.
According to the definition of black hole horizon we have $g(r_H)=0$ and then, 
in order to avoid unphysical divergent quantities we must impose $A(r_H)=B(r_H)=0$ \cite{Horowitz:2010gk}.
Smoothness of solutions at the horizon implies a further consistency constraint \cite{Horowitz:2010gk}:
multiplying the scalar equations by $g(r)$ and then considering them at $r=r_H$ one
obtains that the value of a scalar field and its derivative at $r_H$ are not independent quantities.

All in all, we have a four-parameter family of consistent solutions; as the consistency requirements
impose some conditions at the horizon, our numerical code is build to propagate solutions from 
the horizon towards the boundary. The four ``free'' inputs are therefore the following horizon
quantities,
\begin{equation}\label{fp}
 \phi'(r_h)\ , \ \ \ \
 v'(r_h)\ , \ \ \ \
 \lambda(r_h)\ , \ \ \ \psi(r_h)\ .
\end{equation}

Now we turn the attention to some physical features.
In our analysis we always require that either the leading or the subleading term in the UV expansion of 
both bulk scalars is set to zero. We choose $m^2 = -2$ for both scalars, this implies that we fall
inside the window where two quantizations are allowed \cite{Klebanov:1999tb}; on practical grounds
this means that we have the possibility of interpreting either the leading fall-off as the source
and the subleading as the VEV of the corresponding dual operator or vice versa.
Even without invoking any more sophisticated argument, requiring zero scalar sources 
corresponds to ask for unsourced operators VEV's, i.e. spontaneous condensations.
Furthermore, we decide to work in the canonical ensemble and therefore the charge 
densities (i.e. the subleading terms in the UV behavior of the gauge fields, see Eq. (\ref{UV})) are imposed from outside.
The four parameters (\ref{fp}) are therefore fixed and the code implements such physical inputs
by means of a shooting method%
\footnote{Some further technical detail is given in Section (\ref{tcode}).}.

An important observation is in order:
As already noted, because of the scaling (\ref{scaling}), the physically meaningful
quantities need to be scaling invariant. When solving the holographic system we want to study
a specified thermodynamical state $(\rho_A,\rho_B,T)$ but we are insensitive to an underlying
scaling factor acting as in (\ref{scaling}). Such feature can be read as emerging from the underlying conformal structure
of the strongly coupled quantum field theory.

We define the following ``physical'' (i.e. scaling invariant) quantities
\begin{equation}
 \tilde{T} = \frac{T}{\sqrt{\rho_A+\rho_B}}\ , \ \ \ \ \ \
 \tilde{\cal O}_x = \frac{{\cal O}_x}{(\sqrt{\rho_A+\rho_B})^x}\ , \ \ \ \ \ \
  \tilde{\cal P}_x = \frac{{\cal P}_x}{(\sqrt{\rho_A+\rho_B})^x}\ , 
\end{equation}
where $x=1,2$ and the condensates ${\cal O}_x$ and ${\cal P}_x$ have been defined in (\ref{sates}).

\section{Coexistence and mutual competition/enhancement of the two condensates}

We want to study how the presence of the $\lambda$-condensate
affects the condensation of $\psi$; in particular, we are interested
in how $\gamma$, the direct coupling  between $\psi$ and $\lambda$,
influences the mutual competition/enhancement of the two order parameters.
We plot the $\psi$-condensate as a function of the temperature $\tilde{T}$
in the canonical ensemble, namely keeping the charge
densities $\rho_A$ and $\rho_B$ fixed, in various situations:
with and without the $\lambda$-condensate and for different values of $\gamma$
(of either signs). In Figure \ref{COEX} we show two examples of coexisting $\psi$ 
and $\lambda$ condensates.

\begin{figure}[t]
\centering
\includegraphics[width=70mm]{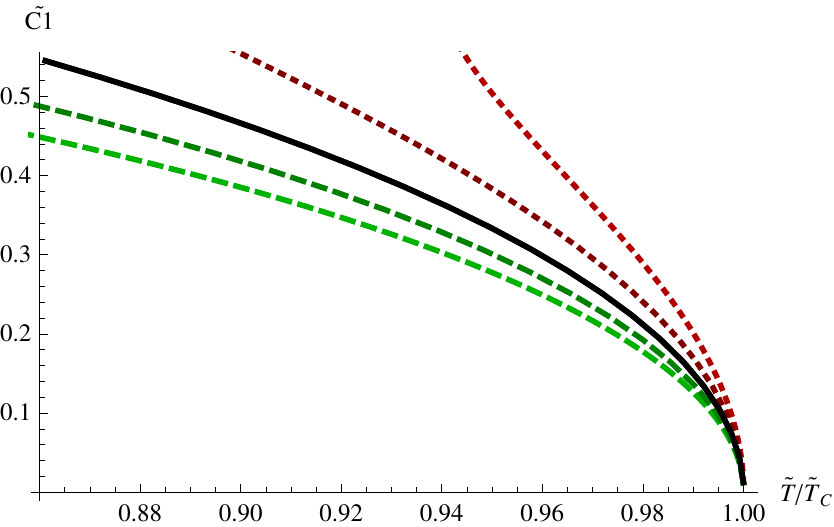} \hspace{0.5cm}
\includegraphics[width=70mm]{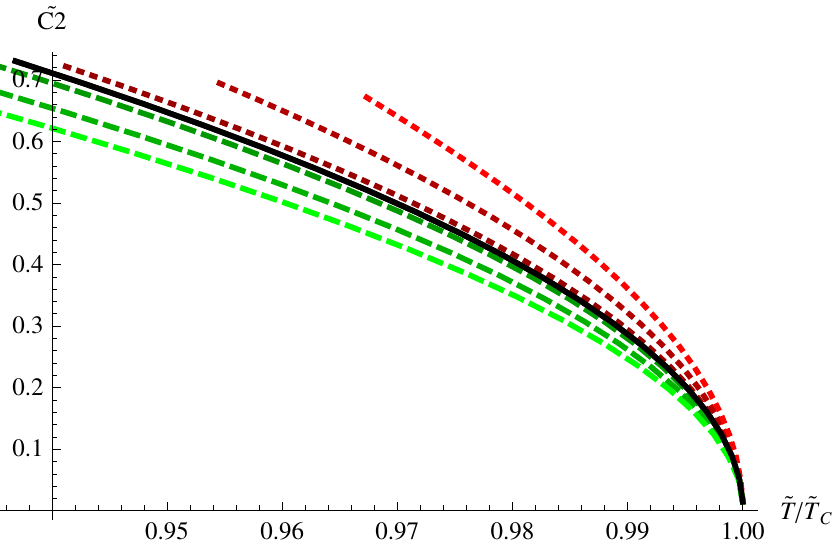}
\caption{The plots correspond to the two possible $\psi$-condensates in the 
balanced $\rho_A=\rho_B$ case. The solid black line represent the non interacting setting ($\gamma = 0$);
red dotted lines correspond to increasingly negative (repulsive) values of $\gamma$ while green dashed
lines refer to increasingly positive (attractive) values of $\gamma$. The $\lambda$-condensate
is not plotted. We have competition/enhancement
corresponding respectively to attractive/repulsive interaction. 
In the balanced case the critical temperature is not affected.}
\label{Probila}
\end{figure}

In Figure \ref{Probila} we show the effects of the $\gamma$ interaction
on the plotted $\psi$-condensates (for both the two different quantization schemes)
in a balanced case, namely $\rho_A=\rho_B$. 
The black solid line represents the $\gamma=0$ case where the two condensates
are completely independent; in this case the presence of a $\lambda$-condensate
has no effect on the $\psi$-condensate. The green dashed condensates correspond
to increasing attractive%
\footnote{The attractive/repulsive character of the $\gamma$ interaction
can be checked comparing to the case in which the interactions between $\psi$ and $\lambda$
are mediated by a would-be scalar (i.e. with even spin) particle.}
(i.e. $\gamma>0$) $\gamma$-couplings while the red dotted
condensates correspond to increasingly repulsive interaction. 
In this balanced case the system is symmetric under the exchange of the ``electric''
and ``magnetic'' sectors and the $\lambda$-condensates (not plotted) are identical
to their $\psi$ cousins. Attractive values of $\gamma$ disfavor the condensates
attenuating their amplitudes while negative values of $\gamma$ have the opposite effect.
Notice that in this balanced case there is no effect on the critical temperature at which
condensation occurs.

\begin{figure}[t]
\centering
\includegraphics[width=70mm]{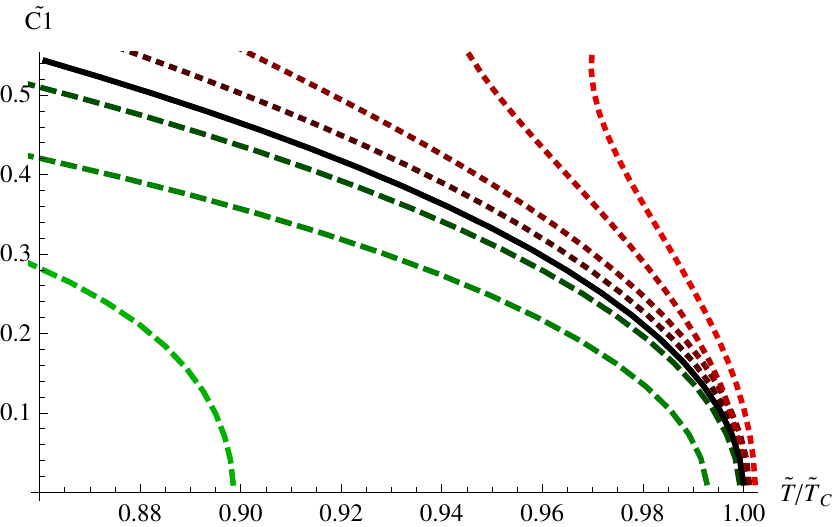} \hspace{0.5cm}
\includegraphics[width=70mm]{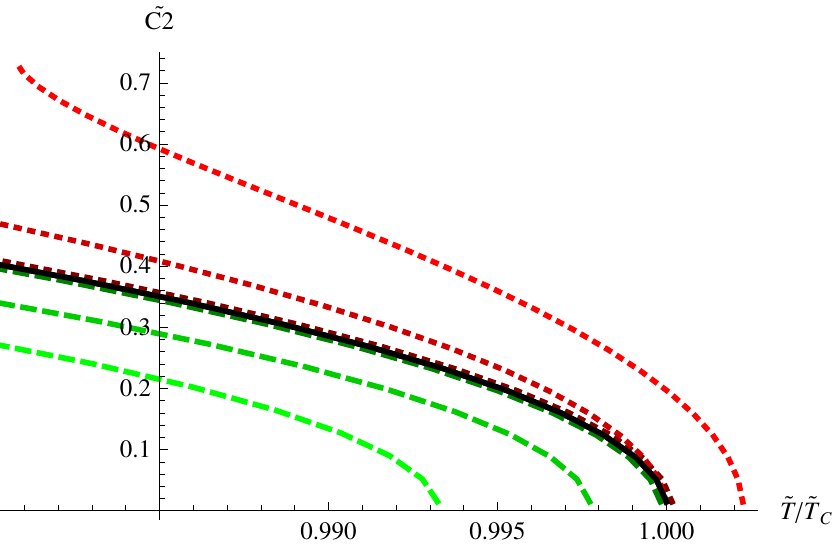}
\caption{Plots showing the two $\psi$-condensates in the unbalanced case. We observe again
(as in the balanced cases plotted in Figure \ref{Probila})
competition/enhancement for attractive/repulsive interaction but here we have
in addition that the critical temperature is accordingly affected.
The plots are traced for $\rho_A\neq \rho_B$ with $\rho_A$ and $\rho_B$
kept fixed; the temperature is normalized with respect to $\tilde{T}_C$, i.e.
the critical temperature for the $\gamma=0$, non-interacting case.}
\label{Probe}
\end{figure}

Turning to an unbalanced case, $\rho_A \neq \rho_B$, we still have the shrinking/amplifying
pattern already occurring in the balanced case but, in addition, also the critical temperature
at which $\psi$ condenses is affected. In particular, attractive interactions lower the critical temperature
while repulsive interactions raise it. This is in line with the idea of competition/enhancement already 
suggested in the balanced case. Indeed a higher value of the critical temperature corresponds to an easier condensation.
We show our results for the unbalanced case in Figure \ref{Probe}.

In \cite{Basu:2010fa} it was conjectured that two holographic orders are competing or
mutually enhancing when their interaction is respectively attractive or repulsive.
The present model offers then a neat example where this conjecture is confirmed.
In Eq. (\ref{psip}) and (\ref{lambdap}) the mixing terms containing $\gamma$ could be thought of
as contributing to the effective mass of the scalars. As a consequence we have for instance that
for $\psi$ the presence of a non -vanishing $\lambda$ affects its effective mass.
We furthermore see that a negative $\gamma$ lowers the effective mass while a positive
$\gamma$ has the opposite effect. In relation to the BF bound criterion for scalar 
instabilities we know that a lower mass value encourages the condensation (see Subsection \ref{stab}).

\subsection{Comments on stability}
\label{stab}
In \cite{Bigazzi:2011ak}, studying the unbalanced holographic superconductor
with a single order parameter, the criterion for IR stability of a scalar field on
the doubly charged $AdS$-Reissner-Nordstr\"{o}rm black hole was given%
\footnote{See \cite{Horowitz:2009ij} for the analogous study in the balanced case.}.
The supporting argument is based on the analysis of scalar fluctuations on the Reissner-Nordstr\"{o}rm background 
in the near-horizon region; there the scalar equation turns out to be that of a free 
scalar on $AdS_2$ geometry with appropriate effective mass. Eventually, the $AdS_2$ 
Breitenlohner-Freedman bound gives the criterion for IR stability of the scalar.
An analogous approach appears to be not feasible here. In principle a similar near-horizon 
analysis could be performed, but here we encounter a problem: actually we would like
to study one scalar on a hairy black hole background (namely on a solution where the 
other, say $\lambda$, scalar has already undergone condensation); even though we have an analytical
expression for the fields of the hairy black hole in the IR region \cite{Horowitz:2009ij}, 
on such background the equation for $\psi$ fluctuations does not reduce to a simple free scalar on an
$AdS$ spacetime. In other words, here we cannot just apply a simple BF bound argument to study
IR stability.

Another aspect related to the ``stability'' of our solutions in a broader sense concerns 
the numerical approach.
In fact, the numerical computations performed to study the thermodynamics of the holographic system
at hand are rather delicate and sensitive. Solutions with coexisting $\psi$ and $\lambda$ 
condensates are sometimes numerically unstable. An excessively low numerical accuracy can 
lead to unphysical ``transitions'' between a double condensate phase to a single condensate 
phase along the numerical sweep in temperature.
A numerical instability must however not be confused with a thermodynamical one.
Indeed, in the holographic framework, the numerical computations consist essentially in
the numerical solution of a system of coupled differential equations, i.e. the classical
equations of motion of the dual bulk gravity theory. Said otherwise, we are not performing any
direct thermodynamical sampling of the partition function of the boundary theory. 

The thermodynamical stability or metastability of the coexisting phase can be checked explicitly
studying the value of the renormalized Euclidean on-shell action of the corresponding dual solutions. For the 
sake of simplicity we have checked successfully the case where the scalar condensates have dimension $2$
(i.e. we chose the UV boundary conditions which set the fastest fall off to zero for both
the scalars); in this circumstance the on-shell Euclidean action is finite \cite{Hertog:2004rz}.


Throughout this paper we have considered both signs for the quartic coupling $\gamma$.
One could worry about the stability of the scalar potential and particularly about
the fact that it could be unbounded from below. However, we work in a probe approximation
where the fields are regarded as small fluctuations; consequently, our effective model
can be thought of to be corrected by higher order terms in the potential which nevertheless
are negligible as long as the scalar fields are small fluctuation.

\subsection{A technical comment on the code}
\label{tcode}
In the first model of a probe holographic superconductor \cite{Hartnoll:2008vx},
the code employed kept the horizon radius (and consequently the temperature) fixed.
The sweep in temperature was actually performed varying $\psi(r_H)$ and letting $\mu$
free; in fact in  \cite{Hartnoll:2008vx} the condensate is plotted with respect to the ratio $T/\mu$ and, 
taking advantage of a scaling analogous to (\ref{scaling}), the plot is interpreted as a
sweep at fixed $\mu$ and progressively lower temperature.
We adopt a similar approach here. We fix the horizon radius $r_H$ to one and require
to the shooting procedure to keep fixed the ratio between the two charge densities.
In this way the plotted points corresponding to various values of the ratio $T/\sqrt{\rho_A+\rho_B}$
can be interpreted as a sweep performed varying the temperature and keeping the charge
densities fixed.

\section{Conclusion and future developments}

We introduced a simple holographic model presenting the coexistence
of two concomitant, non-trivial order parameters. The model represents a simple
generalization of the standard holographic superconductor. Our focus was on
the thermodynamical equilibrium properties of the system with particular 
attention on the mutual interaction of the two orderings.
Contrarily to the general weak-coupling expectation, multiple orderings at strong coupling
can coexist and even enhance one another. We observed the possibility of such an enhancement and,
relying on a probe analysis, we found results in accordance with the conjecture advanced in \cite{Basu:2010fa}
claiming that an attractive/repulsive interaction between the dual bulk fields
leads to competition/enhancement of the corresponding order parameters.
Our model possesses two charge densities associated to two Abelian symmetries $U(1)_A$, $U(1)_B$;
working in the canonical ensemble we studied the coexistence of two orderings
both in an unbalanced ($\rho_A\neq\rho_B$) and in a balanced ($\rho_A=\rho_B$) setting.

Ongoing analysis of the backreacted case indicates that
in general, since it is an attractive interaction, gravity tends to render the
two condensations competitive. However, still on a preliminary level,
the situation could be more complicated: gravitational interactions
are indeed not effectively accountable with a simple shift in the effective mass of the two scalars
as it happened for the direct $\gamma$ interaction in the probe analysis.

One future point of interest consists in the study of transport properties
of the unbalanced, backreacted holographic superconductor with two order parameters.
According to \cite{Bigazzi:2011ak}, in the normal phase, the linear response
of the system to external perturbations could be accounted for by a single frequency dependent 
``mobility'' function. However, in the condensed or doubly condensed phases,
the expectation is that possibly the introduction of a second mobility function 
could account for the transport properties of the ``normal'' and ``condensed'' 
part of the holographic strongly coupled system. The system introduced in the present paper could offer
non trivial checks of this expectation.

Another ongoing interesting perspective (both in relation to condensed matter and QCD) 
is to consider $\lambda$ to be charged
under both gauge fields. In the QCD perspective this brings our model closer to model a
quark-quark condensate and a chiral condensate. In the superconductor sense, instead,
this situation could represent the coexistence of two superconducting condensates, one s-wave
and the other effectively mimicking a p-wave order parameter; said otherwise, we could possibly
have two coexisting Cooper-like pairings both in singlet and triplet spin state.
Still on a rather speculative level, it could be interesting to further study whether 
the present model could furnish a simplified environment to study a bi-layered holographic 
superconductor where the two order parameters would represents both superconducting order 
parameters associated to two different layers.

\section{Acknowledgments}

I am specially grateful to Francesco Bigazzi who suggested me the present study.
I would like to thank also Aldo Cotrone, Andrea Mezzalira, Jan Zaanen, Nicola Maggiore, Nicodemo Magnoli, Alessandro Braggio,
Andrea Amoretti, Daniel Arean, Riccardo Argurio, Diego Redigolo, Katherine Michele Deck, Lorenzo Calibbi, Christoffer Petersson 
and Giorgio Musso for extremely useful discussions and suggestions.

\appendix

\end{document}